\documentclass[aps,prd, nofootinbib,preprintnumbers]{revtex4-2}%
\usepackage{amsmath,amsfonts,amssymb,graphicx,graphics,color, hyperref}
\usepackage[latin9]{inputenc}
\usepackage{pifont}
\usepackage{natbib}
\usepackage{subfigure,epsfig,epstopdf}
\usepackage{bm}
\usepackage{enumerate}
\usepackage{amsmath}
\usepackage{amsfonts}
\usepackage{amssymb}
\usepackage{graphicx}
\usepackage{orcidlink}%
\providecommand{\U}[1]{\protect\rule{.1in}{.1in}}

\newcommand{\be}{\begin{equation}}
\newcommand{\ee}{\end{equation}}
\newcommand{\bea}{\begin{eqnarray}}
\newcommand{\eea}{\end{eqnarray}}

\begin{document}
\title{Inhomogeneous Cosmologies with Bumpy Structure from Superfluid Pionic Vortices
and the propagation of electromagnetic field}
\author{Fabrizio Canfora}
\email{fabrizio.canfora@uss.cl}
\affiliation{Centro de Estudios Científicos (CECs), Casilla 1469, Valdivia, Chile}
\affiliation{Facultad de Ingeniería, Universidad San Sebasti\'an, sede Valdivia, General
Lagos 1163, Valdivia 5110693, Chile}
\author{Alex Giacomini}
\email{alexgiacomini@uach.cl}
\affiliation{Instituto de Ciencias Fisicas y Matem\'aticas, Universidad Austral de Chile,
Valdivia, Chile}
\author{Nikolaos Dimakis,\orcidlink{0000-0002-1307-1073}}
\email{nikolaos.dimakis@ufrontera.cl}
\affiliation{Departamento de Ciencias Físicas, Universidad de La Frontera, Casilla 54-D,
4811186 Temuco, Chile}
\author{Andronikos Paliathanasis,\orcidlink{0000-0002-9966-5517}}
\email{anpaliat@phys.uoa.gr}
\affiliation{Institute of Systems Science, Durban University of Technology, Durban 4000,
South Africa.}
\affiliation{Centro de Investigaci\'on, Innovaci\'on y Creaci\'on (CIIC), Universidad
Cat\'olica de Temuco, Temuco, Chile}
\affiliation{Departamento de Ciencias Matemáticas y Físicas, Facultad de Ingeniería,
Universidad Cat\'olica de Temuco, Temuco, Chile}
\affiliation{National Institute for Theoretical and Computational Sciences (NITheCS), South Africa.}

\begin{abstract}
In this work we introduce exact cosmological solutions in General Relativity
in (3+1) dimensions which take into account both the usual fluid component as
well as a superfluid component describing superfluid Pionic vortices. Such
superfluid vortices are source of inhomogeneities which can be treated
exactly, no perturbation theory is required. These analytic solutions are
relevant to obtain a non-perturbative description of inhomogeneities in
cosmology. We discuss the peculiar effects of these hadronic inhomogeneities
on the propagation of electromagnetic fields on such cosmological backgrounds
analyzing possible observable effects. Finally, the properties that the
effective superfluid satisfy for such solutions to exist are discussed,
together with possible extensions of this family of solutions.

\end{abstract}
\maketitle
\tableofcontents

\affiliation{Centro de Estudios Científicos (CECs), Casilla 1469, Valdivia, Chile}
\affiliation{Facultad de Ingeniería, Universidad San Sebasti\'an, sede Valdivia, General
Lagos 1163, Valdivia 5110693, Chile}

\affiliation{Instituto de Ciencias Fisicas y Matem\'aticas, Universidad Austral de Chile,
Valdivia, Chile}

\affiliation{Departamento de Ciencias Físicas, Universidad de La Frontera, Casilla 54-D,
4811186 Temuco, Chile}

\affiliation{Institute of Systems Science, Durban University of Technology, Durban 4000,
South Africa.}
\affiliation{Centro de Investigaci\'on, Innovaci\'on y Creaci\'on (CIIC), Universidad
Cat\'olica de Temuco, Temuco, Chile}
\affiliation{Departamento de Ciencias Matemáticas y Físicas, Facultad de Ingeniería,
Universidad Cat\'olica de Temuco, Temuco, Chile} \affiliation{National Institute for Theoretical and Computational Sciences (NITheCS), South Africa.}

\affiliation{Centro de Estudios Científicos (CECs), Casilla 1469, Valdivia, Chile}
\affiliation{Facultad de Ingeniería, Universidad San Sebasti\'an, sede Valdivia, General
Lagos 1163, Valdivia 5110693, Chile}

\affiliation{Instituto de Ciencias Fisicas y Matem\'aticas, Universidad Austral de Chile,
Valdivia, Chile}

\affiliation{Departamento de Ciencias Físicas, Universidad de La Frontera, Casilla 54-D,
4811186 Temuco, Chile}

\affiliation{Institute of Systems Science, Durban University of Technology, Durban 4000,
South Africa.}
\affiliation{Centro de Investigaci\'on, Innovaci\'on y Creaci\'on (CIIC), Universidad
Cat\'olica de Temuco, Temuco, Chile}
\affiliation{Departamento de Ciencias Matemáticas y Físicas, Facultad de Ingeniería,
Universidad Cat\'olica de Temuco, Temuco, Chile} \affiliation{National Institute for Theoretical and Computational Sciences (NITheCS), South Africa.}

\section{Introduction}

There is no doubt that the most relevant open problems in physics are
non-perturbative in nature. For instance, the quantum chromodynamics (QCD)
phase diagram is largely determined by topological solitons (see
\cite{R0,R11,R2,Pisarski1,Ref4,Ref5,Ref6,WeinbergBook,BaMa} and references
therein). Similarly, for Bardeen--Cooper--Schrieffer (BCS) superconductors,
vortices play a fundamental role. The main tool to study them is the theory of
Bogomol'nyi--Prasad--Sommerfield (BPS) bounds for the (free) energy of the
configurations in terms of the relevant topological charges. The
(topologically stable) configurations saturating these bounds (BPS solitons
henceforth) are fundamental to explain the non-perturbative features of the
corresponding theories. The BPS bounds are saturated when suitable first order
non-linear differential equations (BPS equations henceforth) are satisfied. It
is worth emphasizing that the BPS equations are far simpler than the second
order field equations. Moreover, it is easy to see that when the BPS equations
are satisfied the second order field equations are satisfied too. Furthermore,
the availability of a saturable BPS bound allows to derive many powerful exact
results, even when the corresponding BPS equations cannot be solved
analytically. In particular, this framework provides a detailed description
both of the low energy dynamics of BPS solitons (see
\cite{moduli,moduli2,moduli3} and references therein) and of the interactions
with Fermions \cite{WeinbergBook}. Thus, these techniques based on BPS
solitons are very effective to analyze both static and dynamical features of
these non-perturbative configurations giving rise to explicit results
difficult to obtain with different techniques.

Due to the fundamental importance of such topological defects, a natural
question to ask is whether or not they leave characteristic fingerprints when
coupled to General Relativity (GR). As far as black hole configurations are
concerned, in vacuum GR the first analytic (3+1)-dimensional examples of black
holes with \textquotedblleft Bumpy event horizons\textquotedblright\ with
matter fields arising from the standard model (Pions) have been constructed in
\cite{BumpyL0,BumpyL1,BumpyL2,BumpyL3,BumpyL4,BumpyL5,BumpyL6,BumpyL7,BumpyL8}. In
these references, the inhomogeneous nature of the event horizon is linked to
the superfluid Pionic vortices, thus avoiding the well known rigidity theorems
in vacuum GR (which restrict both the topology and the curvature of the
horizon: see \cite{N1,N2,N3,N4,N5} and references therein) with a physically
realistic matter field\footnote{Black holes with bumpy horizons are important
in the transient dynamics in black hole mergers as well as in the dynamical
analysis of black strings (see \cite{N18,N19,N20}). The gravitational effects
of bumpy black holes in the analysis of deviations from GR induced by scalar
and vector fields (see \cite{N21,N22,N23,N24,N25,N26} and references therein)
have also been considered.}.

On the other hand, black holes are not, of course, the only physically
relevant configurations in GR. In the present manuscript we will be interested
in dynamical configurations of cosmological type. The fundamental physical
issue we are interested in is a proper non-perturbative description of
inhomogeneities in cosmology. By non-perturbative we mean a description which
does not necessarily assumes that inhomogeneities are ``tiny''. This is a
mandatory step in physical cosmology as, in many situations of interest,
perturbation theory is unsuitable due to the strong fields involved. Moreover,
in cosmology, a normal fluid component is always present to describe the
cosmic dynamics, while the effects of the superfluid components are often
neglected. We will here discuss how one can take into account the
inhomogeneities produced by superfluid vortices together with possible
experimental fingerprints.

Exact solutions of Einstein's field equations that violate the cosmological
principle are of special interest in cosmological studies \cite{ellis}. The
Friedmann-Lemaître-Robertson-Walker (FLRW) geometry, which is the standard
framework in cosmology, is defined by a three-dimensional maximally symmetric
hypersurface, encoding the homogeneity and isotropy of the universe in
accordance with the cosmological principle. In contrast, inhomogeneous
space-times admit fewer or no symmetries, and thus describe more generic
solutions of GR. Inhomogeneous exact space-time solutions allow us to extract
analytic information regarding different epochs of cosmic history and the
properties related to the geometric structure of the universe. Inhomogeneities
could have played an important role in the cosmic evolution; in particular,
small inhomogeneities in the early universe are associated with the small
anisotropies observed in the cosmic microwave background (CMB)
\cite{tur1,tur2}. The geometric richness of inhomogeneous space-times is
further reflected in the diversity of their singular structures, which may
include isotropic, cigar, pancake, oscillatory, and non-scalar singularities
\cite{szek,kras}. Furthermore, exact inhomogeneous solutions provide an
analytic non-perturbative approach, in contrast to standard cosmological
perturbation theory \cite{pp1}.

We will analyze the formation of these very intriguing dynamical gravitating
bumpy configurations in a very conservative cosmological setting. As matter
fields sources for the Einstein equations we will only include a normal fluid
together with the Non-Linear Sigma Model (NLSM) with $SU(2)$ internal symmetry
group. As is well known, this theory describes the low energy dynamics of
Pions and is therefore relevant in all astrophysical and cosmological
situations where hadrons cannot be neglected (see
\cite{R0,R11,R2,Pisarski1,Ref4,Ref5,Ref6,WeinbergBook,BaMa} and references
therein). We will consider configurations of the NLSM representing superfluid
Pionic vortices. On the other hand, as it has been already emphasized, in
order to achieve realistic descriptions of cosmological space-times, it is
necessary to include both a normal fluid component and a superfluid component
(which is often neglected in cosmological settings). Our results show that the
superfluid component can play a prominent role in describing inhomogeneities.

We will adapt the recent results in \cite{US7,US7.5,analytic1,analytic1.5},
where it has been shown how to apply theoretical approaches based on BPS
solitons in the case of superfluids (see for instance,
\cite{GPE1,GPE2,GPE3,GPE4,GPE5,GPE6} and references therein). We will
construct generalizations of the results in
\cite{BumpyL1,BumpyL2,BumpyL3,BumpyL4,BumpyL5,BumpyL6,BumpyL7,BumpyL8} within
the framework of inhomogeneous cosmologies by determining exact Szekeres-like
bumpy space-times. Finally, by using the 1+1+2 decomposition, for the
effective energy momentum tensor for the Pionic source, we demonstrate that we
can generalize known-solutions in less-symmetric geometries.

This paper is organized as follows: In Section \ref{sec-2} we present the
basic elements for the $SU\left(  2\right)  $ NLSM. In Section \ref{sec3} we
review previous results on the existence of exact geometers with a bumpy
configuration. We extend this framework in the case of inhomogeneous
cosmological models, and in Section \ref{sec4} we determine new bumpy
Szekeres-like inhomogeneous cosmological models. Finally, in Section
\ref{sec5} we explore the physical consequences of these hadronic
inhomogeneities on the propagation of electromagnetic fields and in the last
section we draw our conclusions.


\section{BPS equation for vortices in SU(2) NLSM and generalization}

\label{sec-2}

Chiral perturbation theory describes QCD at low energies (see
\cite{R0,R11,R2,Pisarski1,Ref4,Ref5,Ref6,WeinbergBook,BaMa} and references
therein). At leading order the corresponding action is just the Non-Linear
Sigma Model described by the action $S_{NLSM}$. We neglect the subleading
orders in the present work and we consider the $SU(2)$ case. The action is:%
\begin{equation}
S_{NLSM}=\frac{K_{1}}{4}\int d^{4}x\sqrt{-g} \, tr\left(  U^{-1}\partial_{\mu
}U\right)  ^{2}\ , \label{ActionNLSM1}%
\end{equation}
where $f_{\pi}=2\sqrt{K_{1}}$ ($f_{\pi}$ being the Pions decay constant:
$f_{\pi}\approx141\ MeV$).

Without loss of generality, any $SU(2)$-valued scalar field can be represented
as follows:%
\begin{equation}
U=\boldsymbol{1}_{2\times2}\cos\alpha+\sin\alpha n_{j}\mathbf{\tau}^{j}\ ,
\label{parametrization1}%
\end{equation}
where the Isospin vector $n_{j}$ reads%
\begin{equation}
\overrightarrow{n}=(\sin F\cos G,\sin F\sin G,\cos F)\ ,
\label{parametrization2}%
\end{equation}
in which the three scalar degrees of freedom of $SU(2)$ depend, in principle,
on all four-dimensional coordinates:%
\[
\alpha=\alpha(x^{\mu}),\ F=F(x^{\mu}),\ G=G(x^{\mu})\ .
\]
In terms of the parametrization in Eqs. (\ref{parametrization1}) and
(\ref{parametrization2}) the NLSM action in Eq. (\ref{ActionNLSM1}) and the
corresponding field equations read%
\begin{equation}
S_{NLSM}=-\frac{K_{1}}{2}\int d^{4}x\sqrt{-g}\left\{  \left(  \nabla
\alpha\right)  ^{2}+\sin^{2}\alpha\left[  \left(  \nabla F\right)  ^{2}%
+\sin^{2}F\left(  \nabla G\right)  ^{2}\right]  \right\}  \ ,
\label{ActionNLSM2}%
\end{equation}%
\begin{equation}
-\square\alpha+\frac{\sin(2\alpha)}{2}\left[  \left(  \nabla F\right)
^{2}+\sin^{2}F\left(  \nabla G\right)  ^{2}\right]  =0\ , \label{equ1}%
\end{equation}%
\begin{equation}
-\sin^{2}\alpha\square F-\sin(2\alpha)\left(  \nabla\alpha\cdot\nabla
F\right)  +\sin^{2}\alpha\frac{\sin(2F)}{2}\left(  \nabla G\right)  ^{2}=0\ ,
\label{equ2}%
\end{equation}%
\begin{equation}
-\sin^{2}\alpha\sin^{2}F\square G-\sin(2\alpha)\sin^{2}F\left(  \nabla
\alpha\cdot\nabla G\right)  -\sin^{2}\alpha\sin(2F)\left(  \nabla F\cdot\nabla
G\right)  =0\ , \label{equ3}%
\end{equation}
where%
\begin{align*}
\left(  \nabla X\right)  ^{2}  &  =g^{\mu\nu}\left(  \partial_{\mu}X\right)
\left(  \partial_{\nu}X\right)  \ ,\ \ \left(  \nabla X\cdot\nabla Y\right)
=g^{\mu\nu}\left(  \partial_{\mu}X\right)  \left(  \partial_{\nu}Y\right)
\ ,\\
\nabla_{\mu}X  &  =\partial_{\mu}X\ ,\
\end{align*}
namely, $\nabla_{\mu}X$ denotes the usual partial derivative of the scalars
$X$.

The theory minimally coupled to the Einstein-Hilbert action with a
cosmological constant $\Lambda$
\begin{equation}
S=\int d^{4}x\sqrt{-g}\frac{1}{2\kappa}(\mathcal{R}-2\Lambda)+S_{NLSM}\,,
\end{equation}
leads to the Einstein's equation for the metric
\begin{equation}
G_{\mu\nu}+\Lambda g_{\mu\nu}=\kappa\,T_{\mu\nu}\,, \label{eineq}%
\end{equation}
where $\kappa=8\pi G$ (we work in $c=1$ units), $G_{\mu\nu}$ is the Einstein
tensor, and $T_{\mu\nu}=-\frac{2}{\sqrt{-g}}\frac{\delta S_{NLSM}}{\delta
g^{\mu\nu}}$ the energy-momentum tensor of the theory, which reads:
\begin{align*}
T_{\mu\nu}  &  =K_{1}\left\{  \left(  \nabla_{\mu}\alpha\right)  \left(
\nabla_{\nu}\alpha\right)  +\sin^{2}\alpha\left[  \left(  \nabla_{\mu
}F\right)  \left(  \nabla_{\nu}F\right)  +\sin^{2}F\left(  \nabla_{\mu
}G\right)  \left(  \nabla_{\nu}G\right)  \right]  -\right. \\
&  \left.  \frac{g_{\mu\nu}}{2}\left[  \left(  \nabla\alpha\right)  ^{2}%
+\sin^{2}\alpha\left[  \left(  \nabla F\right)  ^{2}+\sin^{2}F\left(  \nabla
G\right)  ^{2}\right]  \right]  \right\}  \ .
\end{align*}

\section{Superfluid BPS Pionic vortices in flat space-times and Bumpy
configurations.}

\label{sec3}

As is well known (see \cite{GibbonsComtet} and references therein) the
following ansatz describes BPS superfluid Pionic vortices:%
\begin{equation}
\alpha=\alpha(x,y)\ ,\ \ F=\frac{\pi}{2}\ ,\ \ G=G(x,y) \label{ansa1}%
\end{equation}
\begin{align}
\partial_{x}\alpha\pm\sin\alpha\partial_{y}G  &  =0\ ,\label{BPSs1}\\
\partial_{y}\alpha\mp\sin\alpha\partial_{x}G  &  =0\ , \label{BPSs2}%
\end{align}
while the metric is%
\[
ds^{2}=-dt^{2}+dx^{2}+dy^{2}+dz^{2}\ .
\]

One can easily see that any solution of the above first order system is also a
solution of the field equations in Eqs. (\ref{equ1}), (\ref{equ2}) and
(\ref{equ3}). Moreover, the first order BPS system in Eqs. (\ref{BPSs1}) and
(\ref{BPSs2}) can be solved explicitly with the following change of variables:%
\[
\partial Z=\frac{\partial\alpha}{\sin\alpha}\rightarrow Z(\alpha)=\log
\tan(\frac{\alpha}{2})\ .
\]
With the new variable $Z=Z(\alpha)$, the BPS equations in Eqs. (\ref{BPSs1})
and (\ref{BPSs2}) reduce to the Cauchy-Riemann equations%
\begin{align}
\partial_{x}Z\pm\partial_{y}G  &  =0\ ,\label{CR1}\\
\partial_{y}Z\mp\partial_{x}G  &  =0\ . \label{CR2}%
\end{align}
Thus, $Z$ and $G$ are the real and the imaginary part of an analytic function
$W(z)$ of $z=x+iy$. If this analytic function is taken as the sum of
logarithms, the corresponding phase corresponds to the Feynman-Onsager
proposal for the phase of a superfluid wave-function \cite{SFP1,SFP2}.

\subsection{Static bumpy configurations}

Now, a natural question which has been analyzed in
\cite{BumpyL1,BumpyL2,BumpyL3,BumpyL4} is to find interesting metrics with the
property that the BPS equations (\ref{BPSs1}) and (\ref{BPSs2}) still
imply the second order field equations for the NLSM in the given metric.

In the above references it has been discussed a very interesting and relevant
class of metrics:
\begin{equation}
ds^{2}=-f(r)dt^{2}+\frac{dr^{2}}{f(r)}+r^{2}\left\{  \exp P(x,y)\left(
dx^{2}+dy^{2}\right)  \right\}  \ . \label{adapted1}%
\end{equation}
The above metric can represent a black hole with an event horizon which is not of 
constant curvature. Indeed, one can solve the fully coupled system of
equations corresponding to GR minimally coupled to the NLSM, where the ansatz
for the $SU(2)$-valued matter field is in Eq. (\ref{ansa1}) with the metric
Eq. (\ref{adapted1}) by taking%
\begin{align}
f(r)  &  =\gamma-\frac{2m}{r}-\frac{\Lambda}{3}r^{2}\ ,\\
-2\kappa K_{1}\left[  (\partial_{x}\alpha)^{2}+(\partial_{y}\alpha
)^{2}\right]   &  =\left(  \frac{\partial^{2}}{\partial x^{2}}+\frac
{\partial^{2}}{\partial y^{2}}\right)  P+2\gamma\exp(P)\ ,
\end{align}
where $\Lambda$ is the cosmological constant. Thus, one can see that when the
matter field is trivial, the equation for $P$ becomes the usual Liouville
equation which implies the event horizon has, locally, constant curvature. It
is worth emphasizing that $\gamma$ (in absence of superfluid Pions) is the
curvature of the event horizon. On the other hand, when the matter field is
non-trivial, the event horizon is not of constant curvature: one can think at
the solution of the above equation for $P(x,y)$ as a surface with curvature
$\gamma$ with inhomogeneities (``bumps'') introduced by the vortices. In the
present manuscript we will disclose the fundamental role of superfluid
vortices in describing inhomogeneous cosmologies non-perturbatively.

\section{Inhomogeneous Cosmological Solutions}

The inhomogeneous Lemaître-Tolman-Bondi (LTB) universe is a spherically
symmetric solution of Einstein's field equations with an inhomogeneous dust
source, which has been widely used to model cosmic voids and the nonlinear
growth of structure \cite{bol1}. In the homogeneous limit, the LTB model
reduces to the FLRW geometry. The LTB geometry has also been used to address
late-time cosmic expansion without invoking a cosmological constant. These
isotropic inhomogeneous geometries belong to a more general family of
inhomogeneous space-times: the Szekeres geometries \cite{szek}. The latter are
exact solutions of GR with an inhomogeneous pressureless fluid source and no
isometries \cite{apost}. They can be classified into two broad classes:
isotropic and inhomogeneous geometries, to which the LTB models belong, and
inhomogeneous anisotropic models, whose ``isotropization'' limit recovers the
Kantowski-Sachs geometry. Generalizations of the Szekeres geometries
incorporating a cosmological constant and an ideal gas equation of state have
been studied in \cite{sz1,sz2}. In the following we consider the existence of
the superfluid in a Szekeres model and we construct inhomogeneous exact solutions.

\label{sec4} For the starting point, we consider the anisotropic and
inhomogeneous line element \cite{szek}
\begin{equation}
ds^{2}=-dt^{2}+e^{2A\left(  t,r,x,y\right)  }dr^{2}+e^{2B\left(
t,r,x,y\right)  }\left(  dx^{2}+dy^{2}\right)  ,
\end{equation}
and additional to the NLSM with the same ansatz for the $SU(2)$-valued matter
field as before, we consider the pressureless matter source
\begin{equation}
T_{\mu\nu}^{\left(  m\right)  }=\rho_{m}\left(  t,r,x,y\right)  u_{\mu}u_{\nu
},
\end{equation}
where $u_{\mu}$ is the comoving observer $u_{\mu}=\delta_{\mu}^{t}$,~$u^{\mu
}u_{\mu}=-1$. In the absence of the NLSM the gravitational field equations
lead to the Szekeres space-times.

The particular choice,
\begin{equation}
A(t,r,x,y)=A(t,r),\quad B(t,r,x,y)=B(t,r)+P(x,y)\,,
\end{equation}
has a minimal effect to the equations of motion of the pions, which are
satisfied with the help of \eqref{ansa1}, \eqref{BPSs1} and \eqref{BPSs2}
under the condition that $G(x,y)$ is a harmonic function. The gravitation
field equations lead to the relations:
\begin{align}
&  2A_{,tt}+2A_{,t}^{2}+2A_{,t}B_{,t}+e^{-2A}\left(  2A_{,r}B_{,r}-B_{,r}%
^{2}-2B_{,rr}\right)  -B_{,t}^{2}-\mathcal{K}\,e^{-2B}-\Lambda=0\,,\\
&  2B_{,tt}+3B_{,t}^{2}-e^{-2A}B_{,r}^{2}+\mathcal{K}\,e^{-2B}-\Lambda=0\,,\\
&  \kappa\rho_{m}=2e^{-2A}\left(  A_{,r}B_{,r}- B_{,rr}\right)  +
2A_{,t}B_{,t} - 3e^{-2A}B_{,r}^{2}+ B_{,t}^{2} + \mathcal{K}\,e^{-2B} -
\Lambda\,\\
&  \left(  \frac{\partial^{2}}{\partial x^{2}}+\frac{\partial^{2}}{\partial
y^{2}}\right)  P+\mathcal{K}\,e^{2P}=-\kappa K_{1}\left[  (\partial_{x}%
\alpha)^{2}+(\partial_{y}\alpha)^{2}\right]  \,,
\end{align}
where $\mathcal{K}$ denotes a separation constant, together with
\begin{equation}
2B_{,tr}-2B_{,r}\left(  A_{,t}-B_{,t}\right)  =0, \label{einnondiag}%
\end{equation}
which is obtained from the non-diagonal component of \eqref{eineq}. The
equation for $\rho_{m}$ implies that the latter can only be a function of $t$
and $r$, i.e. $\rho_{m}=\rho_{m}(t,r)$.

At this point we may distinguish two cases:

\begin{itemize}
\item $B=B(t)$, which immediately satisfies Eq. \eqref{einnondiag}, or,

\item $B$ has a radial dependence, which through Eq. \eqref{einnondiag}
implies a relation between $A$ and $B$. After integration, we obtain
\begin{equation}
\label{AconB}A(t,r) = B(t,r)+A_{0}(r) + \ln(B_{r}(t,r))\, .
\end{equation}
With the above condition, only a single independent equation for $B(t,r)$
remains which is
\begin{equation}
\label{eqB2}2 B_{,tt} +3 B_{,t}^{2}+e^{-2 B} \left(  \mathcal{K} -e^{-2 A_{0}%
}\right)  -\Lambda=0 .
\end{equation}

\end{itemize}

We will refer to the first case as the class I case with metric
\begin{equation}
ds^{2}=-dt^{2}+e^{2A(t,r) }dr^{2}+e^{2B(t)+2 P(x,y)}\left(  dx^{2}%
+dy^{2}\right)  \, ,
\end{equation}
while the second is the class II case with
\begin{equation}
ds^{2}=-dt^{2}+(e^{A_{0}(r)}\partial_{r}(e^{B(t,r)}))^{2} dr^{2}+e^{2B(t,r)+2
P(x,y)}\left(  dx^{2}+dy^{2}\right)  \, .
\end{equation}
Note that the term $e^{2A_{0}(r)}$ can be absorbed in a redefinition of the
radial variable, $e^{2A_{0}(r)} dr^{2}\rightarrow dr^{2}$, or be used to
eliminate the $r$ dependence from the $g_{rr}$ metric component if $B(t,r)$ is
a separable function of the form $B(t,r) = B_{1}(t)+B_{2}(r)$.

From the above metrics, we determine the two following classes of solutions.

\subsection{Class I:\ Anisotropic}

The anisotropic inhomogeneous space-time%
\begin{equation}
ds^{2}=-dt^{2}+\left(  \mu\left(  t,r\right)  +R\left(  t\right)  \right)
^{2}dr^{2}+R^{2}\left(  t\right)  \left(  e^{2P\left(  x,y\right)  }\left(
dx^{2}+dy^{2}\right)  \right)  . \label{ssf1}%
\end{equation}
where the scale factor $R\left(  t\right)  $ and $\mu\left(  t,r\right)  $
satisfy the field equations with a dust fluid source \cite{kassol1},
\begin{equation}
2\frac{\ddot{R}}{R}+\frac{\dot{R}^{2}}{R^{2}}+\frac{\mathcal{K}}{R^{2}%
}-\Lambda=0\,,
\end{equation}%
\begin{equation}
2\mu_{,tt}+2\frac{\dot{R}}{R}\mu_{,t}-\frac{\dot{R}^{2}}{R^{2}}\mu
-\frac{\mathcal{K}}{R^{2}}\left(  \mu+2R\right)  -\Lambda\,\mu=0\,,
\end{equation}%
\begin{equation}
\frac{\dot{R}^{2}}{R^{2}}\left(  1+\frac{2\mu_{,t}}{3\dot{R}}+\frac{\mu}%
{3R}\right)  -\frac{\Lambda}{3}\left(  1+\frac{\mu}{R}\right)  +\frac
{\mathcal{K}}{3R^{2}}\left(  1+\frac{\mu}{R}\right)  =\frac{\kappa}{3}\left(
1+\frac{\mu}{R}\right)  \rho_{m}\,.
\end{equation}
In the above relations, the dot denotes a total derivative with respect to
time and the function $P\left(  x,y\right)  $ is related to the profile
$\mathcal{D}\left(  x,y\right)  $ for the Pions, that is, $\mathcal{D}=-\kappa
K_{1}\left[  (\partial_{x}\alpha)^{2}+(\partial_{y}\alpha)^{2}\right]  $

The constant $\mathcal{K}$ and the function $P\left(  x,y\right)  $ are
related via the condition%
\begin{equation}
\left(  \frac{\partial^{2}}{\partial x^{2}}+\frac{\partial^{2}}{\partial
y^{2}}\right)  P+\mathcal{K}\,e^{2P}=\mathcal{D}\left(  x,y\right)  ,
\label{sf2}%
\end{equation}
which is similar to the condition determining the curvature of the horizon of
a bumpy black hole. In the limit $\mathcal{D}\left(  x,y\right)  =0$, Eq.
\eqref{sf2} becomes the Gauss-Codazzi equation for a 2-surface of constant
Gauss curvature $\mathcal{K}$. The space-time (\ref{ssf1}) belongs to the
Szekeres class of geometries; in the particular case $\mu=\mu(t)$ and
$\mathcal{K}>0$, it reduces to a Kantowski-Sachs space-time, while for
$\mathcal{K}<0$ it is a Bianchi type III locally rotationally symmetric (LRS)
cosmology \cite{decom1}. Zero curvature, $\mathcal{K}=0$, results into a
Bianchi type I LRS model.

\subsection{Class II: Isotropic}

For the second class of inhomogeneous solutions we adopt the ansatz of a
separable function $B(t,r)=\ln\left(  R(t)C(r)\right)  $, which together with
Eq. \eqref{AconB} leads to the line-element%
\begin{equation}
ds^{2}=-dt^{2}+R(t)^{2}\left[  \left(  e^{A_{0}(r)}\frac{dC}{dr}\right)
^{2}dr^{2}+C\left(  r\right)  ^{2}e^{2P\left(  x,y\right)  }\left(
dx^{2}+dy^{2}\right)  \right]  \,.
\end{equation}
We use the word isotropic in this case in a loose sense, just to underline
that the time dependence in all three spatial directions is the same. Note
from the form of the line-element that we can use diffeomorphisms in the $r$
variable to eliminate one of the two functions depending on it. To this end
let us take $C(r)$ as being the radial variable, setting effectively $C(r)=r$.
Elimination of the radial variable from Eq. \eqref{eqB2} when $\mathcal{K}%
\neq0$ implies
\begin{equation}
A_{0}(r)=\frac{1}{2}\ln\left(  \frac{1}{\mathcal{K}-k\,r^{2}}\right)  \,,
\end{equation}
where $k$ is a constant. This brings the line-element to the form
\begin{equation}
ds^{2}=-dt^{2}+R(t)^{2}\left(  \frac{dr^{2}}{\mathcal{K}-k\,r^{2}}%
+r^{2}e^{2P\left(  x,y\right)  }\left(  dx^{2}+dy^{2}\right)  \right)  .
\end{equation}

The scale factor $R\left(  t\right)  $ satisfies the Friedmann equations
\begin{equation}
2\frac{\ddot{R}}{R}+\frac{\dot{R}^{2}}{R^{2}}+\frac{k}{R^{2}}-\Lambda=0\,,
\end{equation}%
\begin{equation}
\frac{\dot{R}^{2}}{R^{2}}+\frac{k}{R^{2}}-\frac{\Lambda}{3}=\frac{\kappa}%
{3}\rho_{m}\,,
\end{equation}
and $P\left(  x,y\right)  $ is given by the Liouville equation with source
(\ref{sf2}). We remark that when $\mathcal{D}\left(  x,y\right)  =0$, the
isotropic and homogeneous FLRW geometry is recovered for $\mathcal{K}\neq0$.
Without loss of generality the above solutions can be generalized when
pressureless fluid source is replaced by an inhomogeneous ideal gas \cite{sz1}.

The energy density of such configurations is localized, and the corresponding
geometry describes localized bumps of finite amplitude whose profile is fixed
by the topological charge and by the coupling constants of the Lagrangian,
rather than by an arbitrary perturbative parameter. Consequently, these
structures lie outside the reach of a linearized treatment around the
homogeneous cosmological background, and the exact approach adopted here
provides analytic control over a genuinely nonlinear inhomogeneous configuration.

The ansatz to describe the Hadronic superfluid vortices has been carefully
chosen in such a way that the Hadronic effects are only visible within the
two-dimensional Gaussian curvature $\mathcal{R}$ for the background geometry.
This is due to the BPS property of the Pionic energy momentum tensor in the
superfluid ansatz which is not spoiled by the gravitational back-reaction.

With our approach, the contributions of the superfluid Hadronic vortices to
the energy momentum tensor reads
\[
T_{\mu\nu}^{Pions}=\frac{1}{\kappa}
\begin{pmatrix}
\mathcal{D}\left(  x,y\right)  e^{-2B} & 0 & 0 & 0\\
0 & -\mathcal{D}\left(  x,y\right)  e^{2\left(  A-B\right)  } & 0 & 0\\
0 & 0 & 0 & 0\\
0 & 0 & 0 & 0
\end{pmatrix}
\ .
\]

From the above expression, one can derive that the Pions do not affect
directly the time-dependence of the metric, but only the inhomogeneous
functions (generating a non-trivial Gaussian curvature of the geometry
\cite{ff1}). This is the technical reason why one can recover Szekeres-like
solutions with a dust fluid source as additional matter term (besides the
superfluid Hadronic vortices), or the spherically symmetric solutions obtained previously.

\section{Electromagnetic fields propagating in cosmological-hadronic
backgrounds}

\label{sec5}

The aim of the present manuscript is to disclose in a clear way the possible
physical effects in cosmology related to inhomogeneities generated by Hadronic
matter fields without using, as far as possible, perturbation theory. Thus, in
this section, we will discuss the propagation of electromagnetic fields in the
cosmological backgrounds discussed here above where the presence of hadrons
cannot be neglected. In other words, we will analyze situations in which the
background on which electromagnetic fields propagate includes not only the
gravitational field but also hadrons. In the available literature, the
hadronic background is often neglected but, as it will be clear in the
following discussion, this can lead to miss relevant physical effects (a
similar sloppiness would lead to miss the Meissner effect in superconductors):
the issue is that charged Hadrons (such as the electrically charged Pions)
exert a force on the photons which, consequently, will deviate from a null
geodesic motion.

In order to develop some intuition, let us shortly discuss the Abelian--Higgs
model (AHM) to disclose the similarities with situations analyzed in this
manuscript. The AHM action is
\begin{equation}
S_{AHM}=\int\left[  -|D_{\mu}\psi|^{2}-\frac{1}{4}\mathcal{F}_{\mu\nu
}\mathcal{F}^{\mu\nu}+\frac{\lambda}{2}\left(  |\psi|^{2}-v^{2}\right)
^{2}\right]  \sqrt{-g}d^{4}x\,, \label{Lag-Abelian-Higgs}%
\end{equation}%
\[
\mathcal{F}_{\mu\nu}=\partial_{\mu}A_{\nu}-\partial_{\nu}A_{\mu}\ ,
\]
where $A_{\mu}$ is the Maxwell field, $\psi$ is the Higgs field, $e$ the
electric charge, $v$ is its vacuum expectation, $\mathcal{F}_{\mu\nu}$ is the
$U(1)$ field strength and $\lambda$ the Higgs coupling constant. The
gauge-covariant derivative is%
\[
D_{\mu}=\partial_{\mu}-ieA_{\mu}\ .
\]
The Higgs field can also be written in \textquotedblleft polar
form\textquotedblright\
\[
\psi=\rho e^{iS},\ \rho\geq0\ ,
\]
where $\rho$ is the amplitude and $S$\ is the phase. In polar variables, the
AHM action is%
\begin{align*}
S_{AHM}=  &  \int\left[  -\left(  D_{\mu}\rho\right)  ^{2}+\frac{\lambda}%
{2}\left(  \rho^{2}-v^{2}\right)  ^{2}-\rho^{2}\left(  D_{\mu}S\right)
^{2}\right]  \sqrt{-g}d^{4}x\\
&  -\frac{1}{4}\int\mathcal{F}_{\mu\nu}\mathcal{F}^{\mu\nu}\sqrt{-g}d^{4}x
\end{align*}
with the $U(1)$ gauge-covariant derivative acting separately on the amplitude
and the phase becomes
\begin{equation}
D_{\mu}\rho=\partial_{\mu}\rho\,,\quad D_{\mu}S=\partial_{\mu}S-eA_{\mu}\,.
\label{covdev1}%
\end{equation}
The Maxwell field equations arising from the Abelian-Higgs model in these
variables read%
\begin{equation}
\nabla_{\mu}\mathcal{F}^{\mu\nu}=c_{0}\rho^{2}\left(  A^{\nu}-\frac{1}%
{e}\nabla^{\nu}S\right)  =c_{0}\rho^{2}\widehat{A}^{\nu}\ , \label{equMAHM}%
\end{equation}%
\begin{align*}
\widehat{A}_{\nu}  &  =\left(  A_{\nu}-\frac{1}{e}\nabla_{\nu}S\right)  \ ,\\
\partial_{\mu}\widehat{A}_{\nu}-\partial_{\nu}\widehat{A}_{\mu}  &
=\partial_{\mu}A_{\nu}-\partial_{\nu}A_{\mu}+\sum_{j}c_{j}\delta_{j}\ ,
\end{align*}
where the $c_{j}\delta_{j}$ terms are possible $\delta$-like singularities
which, however, in the present case will play no role. The above Maxwell
equations are appropriate when the electromagnetic field is propagating on a
background with a non-vanishing condensate $\psi$ and: if the electromagnetic
field is not too intense, the backreaction of the electromagnetic field on the
condensate can be neglected.

As far as the present paper is concerned, the important point is that the
$U(1)$ covariant derivative acting on the Chiral field $U\in SU(2)$ has an
analogous form:%
\[
D_{\mu}U=\partial_{\mu}U+eA_{\mu}\left[  U,t_{3}\right]
\]
which in terms of the three $SU(2)$ degrees of freedom $\alpha$, $F$ and $G$
read
\begin{equation}
D_{\mu}\alpha=\partial_{\mu}\alpha\,,\quad D_{\mu}F=\partial_{\mu}F\,,\quad
D_{\mu}G=\partial_{\mu}G-2 \, e A_{\mu}\,. \label{covdev2}%
\end{equation}
Thus, $G(x^{\mu})$ is like $S(x^{\mu})$ (namely, the phase of the Higgs
field). On the other hand, $\alpha$ and $F$ are not affected directly by the
coupling with the Maxwell field. Consequently, the action which describes the
propagation of the Maxwell field in a Hadronic-gravitational background is:%
\[
S_{NLSM}=- \frac{K_{1}}{2}\int d^{4}x\sqrt{-g}\left\{  \left(  \nabla
\alpha\right)  ^{2}+\sin^{2}\alpha\left[  \left(  \nabla F\right)  ^{2}%
+\sin^{2}F\left(  DG\right)  ^{2}\right]  \right\}
\]%
\begin{equation}
-\frac{1}{4}\int\mathcal{F}_{\mu\nu}\mathcal{F}^{\mu\nu}\sqrt{-g}d^{4}x\ ,
\label{MaxAction}%
\end{equation}

In many important cosmological applications it is mandatory to analyze the
propagation of electromagnetic fields in the given cosmological background
\cite{EMcosmo0,EMcosmo1,EMcosmo2,EMcosmo3}. In the present situation the
novelty is that the background also includes the Pions which are the sources
of the inhomogeneities. The corresponding Maxwell equations are
\begin{equation}
\nabla_{\mu}\mathcal{F}^{\mu\nu}=4K\sin^{2}\alpha\sin^{2}F\left(  A^{\nu
}-\frac{1}{2 \, e}\nabla^{\nu}G\right)  \, , \label{MaxwellEq}%
\end{equation}
where $K=K_{1} e^{2}$. The above equations are valid as long as the
backreaction of the maxwell field on the cosmological-hadronic background can
be neglected (it is widely believed that this approximation is valid in most
of the applications). One can notice that $\sin^{2}\alpha\sin^{2}F$ plays the
role of the amplitude $\rho^{2}$ of the Higgs field. However, an important
difference with respect to the AHM is that in the present case there is an
upper bound to the effective amplitude (due to the presence of the
trigonometric function associated to the $SU(2)$ Chiral field) while there is
not obvious upper bound to $\rho^{2}$. The metric and hadronic backgrounds
encode the interaction of the gauge field with gravity and with the Pionic background.

In the present case of the solutions described in the previous sections
$\sin^{2}F=1$. Hence, a Maxwell probe field satisfies the following
equations:
\begin{align}
\nabla^{\mu}\mathcal{F}_{\mu\nu}  &  =\left(  4K\sin^{2}\alpha\right)
\widehat{A}_{\nu}\ ,\label{maxfinal}\\
\widehat{A}_{\mu}  &  =\left(  A_{\mu}-\frac{1}{2 \, e }\nabla_{\mu}G\right)
\,,\nonumber
\end{align}
where $\alpha(x,y)$ is the $SU(2)$ profile which solves the BPS equations
introduced in the previous section. A quite dramatic physical effect of the
hadronic inhomogeneities is that the Maxwell equations acquire a non-trivial
space-time dependent mass-like term which is maximal at the top of
inhomogeneities sourced by the topological defects (where $\sin^{2}\alpha
\sim1$) while it is very small far from them. This leads to a further
suppression of the Maxwell field (as it would happen in a superconductor)
which is usually neglected.

\subsection{The simplest non-trivial examples}

The simplest non-trivial background where this effect appears is the
\textit{Bumpy de Sitter metric}: {
\begin{equation}
ds^{2}=-dt^{2}+R(t)^{2}\left(  \frac{dr^{2}}{\mathcal{K}-k\,r^{2}}%
+r^{2}e^{2P\left(  x,y\right)  }\left(  dx^{2}+dy^{2}\right)  \right)  .
\label{BumpydS1}%
\end{equation}
}

As it has been already emphasized, $R\left(  t\right)  $ satisfies{
\[
2\frac{\ddot{R}}{R}+\frac{\dot{R}^{2}}{R^{2}}+\frac{k}{R^{2}}-\Lambda=0\,,
\]%
\[
\frac{\dot{R}^{2}}{R^{2}}+\frac{k}{R^{2}}-\frac{\Lambda}{3}=\frac{\kappa}%
{3}\rho_{m}\,.
\]
} On the other hand $P\left(  x,y\right)  $ is given by the Liouville equation
with source in Eq. (\ref{sf2}). Let us consider a situation in which Eq.
(\ref{sf2}) can be solved explicitly with a non-trivial Pionic profile. In
order to proceed, it is convenient the following change of coordinates in the
$(x,y)$ coordinates%
\begin{equation}
x=\rho\cos\zeta\ ,\ \ y=\rho\sin\zeta\ . \label{exe1}%
\end{equation}
Using these coordinates, the chiral profiles $\alpha$ and $G$ read%
\begin{align}
\alpha(x,y)  &  =2\arctan\left(  \exp H\right)  \ ,\label{exe2}\\
H  &  =\log\rho\ ,\ \ G=-\zeta\ . \label{exe3}%
\end{align}
Moreover, with this source, Liouville equation in Eq. (\ref{sf2}) is solved by
the following $P$:%

\[
\exp(2P)=\delta\frac{4}{\left(  1+\rho^{2}\right)  ^{2}},\ \ \delta
=\frac{1-\kappa K_{1}}{\mathcal{K}}%
\]
The metric corresponds to a pair of vortices (one in the north pole, the other
in the south pole) which generate an angular defect (proportional to $\delta$)
in the two-dimensional metric.

Before going into the analysis of the Maxwell equations in this
specific case, it is worth emphasizing the following general considerations.
In the simplest case of one bump it is possible to show (see the discussion
below Eq. (\ref{finalmetric2})) that the metric actually possesses a spherical
 $SO(3)$ symmetry. The main point of our paper is that even in such
simplest case in which one could think (by only looking at the metric) that
there is spherical symmetry, in fact the spherical symmetry is broken by the
interaction of the Maxwell field with the Pions. This can be seen as follows:
in the Maxwell equations for the Maxwell field propagating in the background
with one BPS vortex (Eq. (\ref{finalMax}) here below) $\rho$ is an
angular (stereographic) coordinate. Thus, if one would neglect the right hand
side of Eq. (\ref{finalMax}) one could indeed conclude that there is a
spherical symmetry as the left hand side of the equation possesses the
symmetries of the metric itself. On the other hand, the left hand side of the
above equation cannot be neglected (as neglecting the right hand side in the
present case would be like neglecting the Meissner effect in superconductors).
It is precisely in this point that one can notice the key observation of our
manuscript. The right hand side of the Maxwell equations in non-trivial
Hadronic backgrounds will generically possess less symmetries than the left
hand side. The reason is strictly related to the generalized hedgehog ansatz
for $SU(2)$ valued scalar field (see \cite{BumpyL0}). In order to
keep alive the topological charge (the vorticity in the present case) this
approach requires that the matter field source of the Einstein equation should
have less symmetries than the metric itself since one should only require that
the energy-momentum tensor must be compatible with the space-time symmetries.
In other words, the strategy of the generalized hedgehog ansatz is equivalent
to consider $SU(2)$ valued matter field which do not possess the
space-time symmetries of the metric but in such a way that the corresponding
energy-momentum tensor does. As the present manuscript shows, this is not just
a technical point since if there are test matter fields (such as the photons
in our case) which couple directly to the $SU(2)$  valued matter field
(which describes charged Hadrons) then such a test field will interact
directly with the $SU(2)$-valued matter field (and not just with its
energy-momentum tensor as the space-time metric does). Consequently, this
interaction will not be invariant under all the symmetry of the space-time
metric: this is the unavoidable origin of the inhomogeneities associated to
the propagation of light in Hadronic backgrounds. Roughly speaking, the
photons perceive directly $U\in SU(2)$ (due to minimal coupling of
Maxwell field with charged fields) and since $U$ has less symmetries
than its own $T_{\mu\nu}$ the Maxwell equations will not be
spherically symmetric even if the space-time metric is. Obviously, in the case
of more than one bumps this is even more true as, in these cases, not even the
space-time metric is spherically symmetric but explicit solutions of the
Liouville equation with sources are not available in the general case. On the
other hand, at least when the bumps are well separated, one can assume that
the effects of the bumps on the propagating test Maxwell field can be
described as the sum of the effects of the individual bumps. It is for these
reasons that the configuration under study is strictly inhomogeneous (despite
the apparent form of metric suggesting otherwise).

Resuming, we are interested in analyzing the Maxwell equations in Eq.
(\ref{maxfinal}) within the background metric%
\begin{equation}
ds^{2}=-dt^{2}+R(t)^{2}\left(  \frac{dr^{2}}{\mathcal{K}-k\,r^{2}}+4\delta
r^{2}\frac{\left(  d\rho^{2}+\rho^{2}d\zeta^{2}\right)  }{\left(  1+\rho
^{2}\right)  ^{2}}\right)  \ , \label{finalmetric}%
\end{equation}
and where the Chiral profiles $\alpha$ and $G$ are defined in Eqs.
(\ref{exe2}) and (\ref{exe3}).

Thus, when we study the equation%
\[
\nabla^{\mu}\mathcal{F}_{\mu\nu}=\left(  4K\sin^{2}\alpha\right)  \widehat
{A}_{\nu}%
\]
the metric in Eq. (\ref{finalmetric}) enters in the left-hand side when
computing $\nabla_{\mu}\mathcal{F}^{\mu\nu}$. On the other hand, the Chiral
profiles $\alpha$ and $G$ enter when computing the right hand side $\left(
4K\sin^{2}\alpha\right)  \widehat{A}_{\nu}$. In particular, in this simplest
non-trivial example, $\sin^{2}\alpha$ only depends on $\rho$. Moreover, we
have%
\[
\sin\left(  2\Theta\right)  =\frac{2\tan\Theta}{1+\tan^{2}\Theta}\ ,
\]%
\[
\sin\alpha=\frac{2\rho}{1+\rho^{2}}\ ,\ \sin^{2}\alpha=\frac{4\rho^{2}%
}{\left(  1+\rho^{2}\right)  ^{2}}\ .
\]
Therefore, we have to solve%
\begin{equation}
\nabla^{\mu}\mathcal{F}_{\mu\nu}=\frac{16K\rho^{2}}{\left(  1+\rho^{2}\right)
^{2}}\widehat{A}_{\nu} \label{finalMax}%
\end{equation}
in the metric in Eq. (\ref{finalmetric}). In general, the above equation
cannot be solved analytically (as, due to the photons-Hadrons interactions,
the equation is not separable). However, some general considerations can be made.

Let us consider a rough analogy of the above Maxwell equation with a
Schrodinger equation (we could consider a situation in which $\widehat{A}%
_{\nu}$\ has only one component). Then, the right hand side of Eq.
(\ref{finalMax}) looks like a repulsive potential. This can be seen by putting
all the terms of the equation on the left hand side and observing that the
term $\frac{16K\rho^{2}}{\left(  1+\rho^{2}\right)  ^{2}}\widehat{A}_{\nu}$
has opposite sign compared with the term $\frac{\partial^{2}A}{\partial
\rho^{2}}$. On the other hand, such repulsive term vanishes for small $\rho$
and for very large $\rho$. This means that there is some barrier at some
intermediate value of $\rho$ (which is determined by the maximum of
$V_{eff}=\frac{16K\rho^{2}}{\left(  1+\rho^{2}\right)  ^{2}}$). Thus, if there
is a small-amplitude propagating wave localized around $\rho=0$ at $t=0$, then
the barrier at the maximum of $V_{eff}$ will tend to keep the wave localized
around $\rho=0$. Taking into account that $\rho$ is a sort of angular
coordinate, this argument shows that such Hadronic inhomogeneities will
generate a very strong angle dependence in the propagation of electromagnetic
wave. The expected signal of these Hadronic inhomogeneities are bright
(angles-dependent) spots in the propagation of electromagnetic waves related
to the position(s) of the inhomogeneities. Note that these spots would not
appear if one would (incorrectly) neglect the Hadrons-photons interaction terms.

As an illustrative example, here we will analyze the following Maxwell field%
\begin{equation}
\widehat{A}_{\mu}=\left(  0,0,0,A\left(  t,r,\rho\right)  \right)  \ ,
\label{ansatz}%
\end{equation}
in other words, we will now consider the case of a gauge potential with a
non-vanishing component only along the $\zeta$-direction. First of all, one
can notice that this ansatz is consistent. Indeed, from Eq. (\ref{finalMax}),
it follows that%
\begin{equation}
\nabla^{\mu}\nabla^{\nu}\mathcal{F}_{\mu\nu}=0\rightarrow\nabla^{\mu}\left(
\frac{16K\rho^{2}}{\left(  1+\rho^{2}\right)  ^{2}}\widehat{A}_{\mu}\right)
=0\ . \label{conscond1}%
\end{equation}
It is easy to see that the ansatz in Eq. (\ref{ansatz}) satisfies the above
condition. Equation \eqref{finalMax} for the ansatz \eqref{ansatz} reads
\begin{equation}%
\begin{split}
\label{bigMax} &  4(1-\kappa K_{1})r^{2}\rho\left(  \rho^{2}+1\right)
^{2}R^{2}\partial_{t,t}A+4r^{2}\rho\left(  \rho^{2}+1\right)  ^{2}(1-\kappa
K_{1})\left(  kr^{2}-\mathcal{K}\right)  \partial_{r,r}A-\mathcal{K}%
\rho\left(  \rho^{2}+1\right)  ^{4}\partial_{\rho,\rho}A\\
&  +4r^{2}\rho\left(  \rho^{2}+1\right)  ^{2}(1-\kappa K_{1})R\dot
{R}\,\partial_{t}A+4kr^{3}\rho\left(  \rho^{2}+1\right)  ^{2}(1-\kappa
K_{1})\partial_{r}A-\mathcal{K}\left(  3\rho^{2}-1\right)  \left(  \rho
^{2}+1\right)  ^{3}\partial_{\rho}A\\
&  +64e^{2}K_{1}r^{2}\rho^{3}(1-\kappa K_{1})R^{2}A=0.
\end{split}
\end{equation}

It may be more intuitive to change the $\rho$ variable in
\eqref{finalmetric} to a polar angle $\theta$ through the transformation
$\theta\rightarrow\rho=\tan\left(  \theta/2\right)  $, where the metric
assumes a form which makes apparent it's invariance under spatial rotations
\begin{equation}
ds^{2}=-dt^{2}+R(t)^{2}\left(  \frac{dr^{2}}{\mathcal{K}-k\,r^{2}}%
+\delta\,r^{2}\left(  d\theta^{2}+\sin^{2}\theta\,d\zeta^{2}\right)  \right)
.\label{finalmetric2}%
\end{equation}
Nevertheless, this is an inhomogeneous space-time as can be evidenced by
looking at the spatial Ricci curvature. The only spatially homogeneous
space-times invariant under $SO(3)$ are the FLRW space-time and the
Kantowski-Sachs geometry, where in both cases ${}^{(3)}\mathcal{R}$ is
constant. For the metric \eqref{finalmetric2} we can calculate
\begin{equation}
{}^{(3)}\mathcal{R}=6k+\frac{2\kappa K_{1}\mathcal{K}}{r^{2}(1-\kappa K_{1})},
\end{equation}
which demonstrate the inhomogeneity being due to the pionic content since
$K_{1}$ and $\mathcal{K}$ are different from zero.

In the $\theta$ variable Maxwell's equation \eqref{bigMax} reads
\begin{equation}
\label{bigMax2}%
\begin{split}
&  r^{2} (1-\kappa K_{1}) R^{2} \partial_{t,t} A +(1- \kappa K_{1}) \left(  k
r^{2}- \mathcal{K}\right)  r^{2} \partial_{r,r} A - \mathcal{K}\,
\partial_{\theta,\theta} A+ r^{2} (1-\kappa K_{1}) R \dot{R} \partial_{t} A +
k r^{3} (1-\kappa K_{1}) \partial_{r} A\\
&  + \mathcal{K} \cot\theta\, \partial_{\theta} A + 4 e^{2} K_{1} (1-\kappa
K_{1}) r^{2} \sin^{2}\theta R^{2} A =0 .
\end{split}
\end{equation}
It is our purpose to study numerical solutions to this equation under certain
reasonable boundary conditions and how they affect measurable quantities. To
this end, let us introduce the energy-momentum tensor of the electromagnetic
field
\begin{equation}
T_{\mu\nu}^{\text{em}} = \frac{1}{\mu_{0}} \left(  F_{\mu\kappa}F_{\nu
}^{\;\;\kappa} - \frac{1}{4} g_{\mu\nu} F_{\kappa\lambda}F^{\kappa\lambda}
\right)  ,
\end{equation}
and the standard definitions of the electric and magnetic field
\begin{subequations}
\label{emfields}%
\begin{align}
E_{\mu}  &  = F_{\mu\nu} u^{\nu},\\
B_{\mu}  &  = \frac{\sqrt{-g}}{2} \epsilon_{\mu\kappa\sigma\tau} u^{\kappa
}F^{\sigma\tau},
\end{align}
respectively, where $\epsilon_{\mu\kappa\sigma\tau}$ is the Levi-Civita tensor
density with $\epsilon_{0123}=+1$. With our assumptions, the expressions for
the fields are
\end{subequations}
\begin{align}
\label{magfield}E_{\mu}  &  = (0,0,0,\partial_{t} A)\\
B_{\mu}  &  = \left(  0, \frac{-\mathcal{K} \partial_{\theta}A}{ (1-\kappa
K_{1}) r^{2} \left(  \mathcal{K}-k r^{2}\right)  ^{\frac{1}{2}} R \sin\theta},
\frac{\left(  \mathcal{K}-k r^{2}\right)  ^{\frac{1}{2}} \partial_{r} A}{R
\sin\theta}, 0 \right)  .
\end{align}

In order to simplify the setting, we shall consider a constant scale
factor for the background, $R(t)=$const., which permits the separation of
variables in equation \eqref{bigMax2} by introducing $A=\exp(\lambda
t)\tilde{A}(r,\theta)$; the resulting equation being
\begin{equation}
r^{2}(1-\kappa K_{1})\left(  K-kr^{2}\right)  \partial_{r,r}\tilde
{A}+\mathcal{K}\partial_{\theta,\theta}\tilde{A}-kr^{3}(1-\kappa
K_{1})\partial_{r}\tilde{A}-\mathcal{K}\cot\theta\,\partial_{\theta}\tilde
{A}-r^{2}R^{2}(1-\kappa K_{1})\left(  4e^{2}K_{1}\sin^{2}\theta+\lambda
^{2}\right)  \tilde{A}=0.\label{bigMax3}%
\end{equation}
We further assume $\lambda=0$, so that we are left with a pure magnetic field.
It is important to note that the expression for the $B_{\mu}$ is singular at
$r=0$ and at $\theta=0$ and $\theta=\pi$. It is thus important to impose
boundary conditions of vanishing $A$ at this points. In what follows we
present some examples.

\subsection{A dipole magnetic moment}

Let us consider a magnetic dipole moment aligned with the $z$-axis and
positioned at the origin. We impose the Dirichlet boundary conditions:
\begin{equation}
\tilde{A}(0,\theta)=\frac{1}{4}\sin^{2}\theta,\quad\tilde{A}(r_{max}%
,\theta)=\tilde{A}(r,0)=\tilde{A}(r,\pi)=0.
\end{equation}
Notice that the $r$ variable is bounded (as also happens in the FLRW case for
$k\neq0$) to $r_{max}=\sqrt{\mathcal{K}/k}$ so that the metric
\eqref{finalmetric2} maintains a Lorentzian signature.

\begin{figure}[h]
\centering
\includegraphics[scale=0.5]{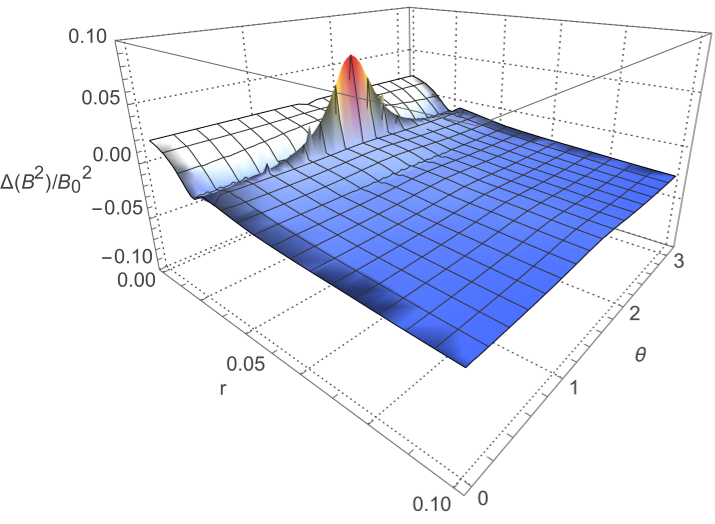}\caption{The relative difference
$\Delta(B^{2})/B_{K_{1}=0}^{2}$, vacuum vs pionic matter content, of the
squares of the amplitudes of the magnetic fields for the case of a dipole
magnetic moment.}%
\label{fig1}%
\end{figure}

\begin{figure}[h]
\centering
\includegraphics[scale=0.5]{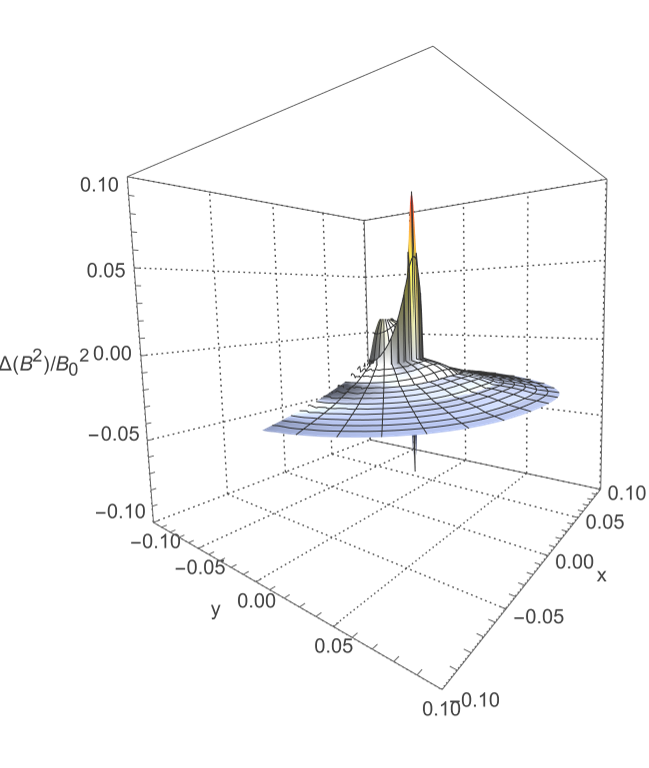}\caption{The relative difference
$\Delta(B^{2})/B_{K_{1}=0}^{2}$, vacuum vs pionic matter content, of the
squares of the amplitudes of the magnetic fields for the case of a dipole
magnetic moment in Cartesian coordinates $x$, $y$. In this graph we
distinguish also a visible negative spike.}%
\label{fig1b}%
\end{figure}

In order to distinguish the interaction, we numerically solve the
equation \eqref{bigMax3}, for $\lambda=0$ with the above boundary conditions
for two distinct cases: a) the pion free case with $K_{1}=0$, and b) the case
$K_{1}\neq0$ which imposes the additional mass term. In each case we calculate
the square of the amplitude of the magnetic field which for our ansatz reads:
\begin{equation}
B^{2}=B_{\mu}B^{\mu}=\frac{K\left(  r^{2}(1-\kappa K_{1})\left(
\mathcal{K}-kr^{2}\right)  (\partial_{r}\tilde{A})^{2}+\mathcal{K}%
(\partial_{\theta}\tilde{A})^{2}\right)  }{r^{4}\sin^{2}\theta(1-\kappa
K_{1})^{2}R^{4}}.
\end{equation}
In order to disclose the physical effects of the Hadrons on the propagation of
light (which is often neglected in the literature), in Fig. \ref{fig1} we have
analyzed the relative difference
\begin{equation}
\frac{\Delta(B^{2})}{B_{K_{1}=0}^{2}}=\frac{B^{2}-B_{K_{1}=0}^{2}}{B_{K_{1}%
=0}^{2}}.\label{relativeB}%
\end{equation}
The plot manifests a finite spike centered at the equator $\theta=\pi/2$ and
to a finite distance from the origin $r=0$, which varies symmetrically with
the polar angle: such a spike is a genuine effect of the photons-Hadrons. For
the numerics, we used the values $k=\mathcal{K}=e=\mu_{0}=R=1$ and
$K_{1}=0.01$. In Fig. \ref{fig1b} we depict the same function in terms of
Cartesian coordinates $x=r\cos\theta$ and $y=r\sin\theta$. 


\subsection{Quadrupole magnetic moment}

\begin{figure}[h]
\centering
\includegraphics[scale=0.5]{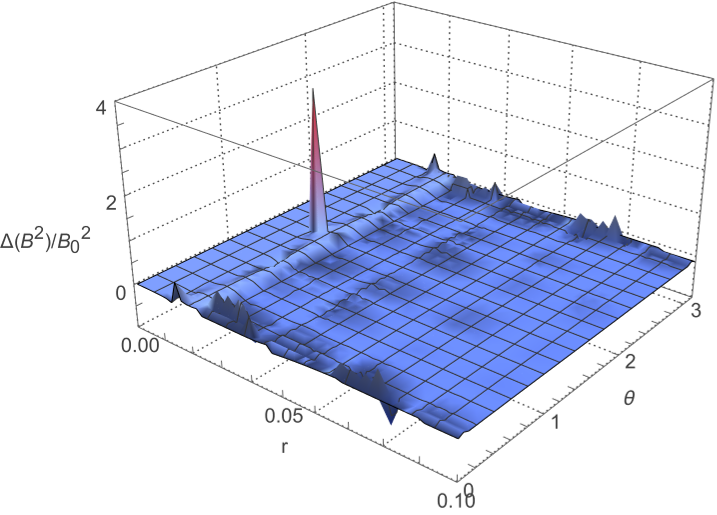}\caption{The relative difference
$\Delta(B^{2})/B_{K_{1}=0}^{2}$, vacuum vs pionic matter content, of the
squares of the amplitudes of the magnetic fields for the case of a quadrupole
magnetic moment. We observe a larger and more central spike than in the case
of the dipole moment. The minor peaks seen close to the edges are numerical
artifacts since the numerical solution confronts the $\theta=0$ and
$\theta=\pi$ points that send to zero the denominator in \eqref{magfield}.}%
\label{fig2}%
\end{figure}

\begin{figure}[h]
\centering
\includegraphics[scale=0.5]{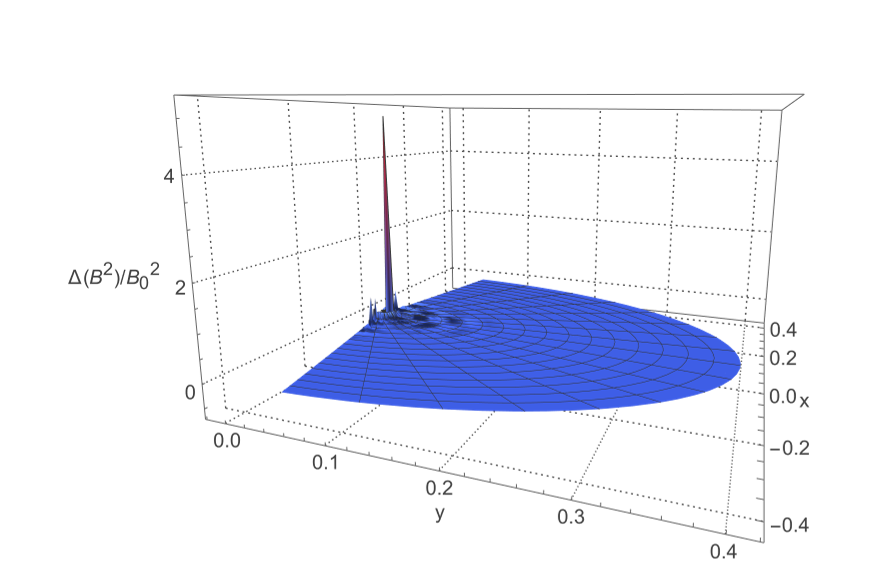}\caption{The relative difference
$\Delta(B^{2})/B_{K_{1}=0}^{2}$, vacuum vs pionic matter content, of the
squares of the amplitudes of the magnetic fields for the case of a quadrupole
magnetic moment in Cartesian coordinates $x$, $y$. }%
\label{fig2b}%
\end{figure}

In this case we consider the same equation under the boundary
conditions
\begin{equation}
\tilde{A}(0,\theta)=\frac{1}{4}\sin^{2}\theta\cos\theta,\quad\tilde{A}%
(r_{max},\theta)=\tilde{A}(r,0)=\tilde{A}(r,\pi)=0,
\end{equation}
corresponding to a quadrupole magnetic moment aligned along the $z$-axis and
placed at the origin. As before we calculate the relative quantity
\eqref{relativeB} for the same set of values $k=\mathcal{K}=e=\mu_{0}=R=1$ and
$K_{1}=0.01$. We see the relative graph in Figs. \ref{fig2} and \ref{fig2b}.
We observe that, despite the different boundary conditions, the spike is very
similar to the previous case. The only difference is that the spike is more
pronounced than before and more centered in the equator $\theta=\pi/2$. 

The two examples discussed here strongly suggest that, independently
on the boundary conditions (as long as these boundary conditions are
physically reasonable), the interactions of the propagating Maxwell field with
the Hadronic vortices will generate bright spots in the amplitude of the
electromagnetic field. It is also useful to remind that this phenomenon that
the Maxwell equations possess less symmetries than the metric itself is
unavoidable when the Hadronic matter is characterized by some non-trivial
topological charge (such as vorticity, Baryonic charge and so on). We will
come back on this very intriguing phenomenon in a future publication. 


\section{Conclusions and perspectives}

\label{sec6}

In the present manuscript, we have constructed analytic cosmological
configurations taking into account both the usual normal fluid component as
well as a superfluid component describing superfluid Pionic vortices. The idea
is that superfluid vortices are source of inhomogeneities and can be treated
analytically, without approximations, thus, no perturbation theory is
required. Consequently, our solutions are relevant to obtain a
non-perturbative description of inhomogeneities in cosmology. This is possible
due to the specific BPS ansatz such that the entire effect of the Hadronic
superfluid is encoded in the source term for the constraint for the
two-dimensional Gaussian curvature. By using this property we determined two
Szekeres-like classes of inhomogeneous cosmological exact solutions.

One of the most interesting effects of these hadronic inhomogeneities
manifests itself in the analysis of the propagation of electromagnetic fields
on these backgrounds. One can show that these Hadronic inhomogeneities induce
electromagnetic waves to ``concentrate'' close to the positions of the
Hadronic vortices generating \textit{bright spots} (namely, local maxima in
the amplitude of the propagating electromagnetic field) which, at least in
principle, should be easy to detect.

These results open many interesting perspectives. First of all, it is worth
trying to push as far as possible a non-perturbative description of (Hadronic)
inhomogeneities in cosmology: the present construction turns out to be quite
flexible in this respect. For instance, the pressureless source may be
replaced by an inhomogeneous ideal gas which will provide inhomogeneous
solutions which belong to the Szekeres-Szafrom geometries. Furthermore, in a
similar way, the cosmological constant can be introduced.

Secondly, it is worth analyzing in more details the observational effects that
these Hadronic inhomogeneities have on the propagation of electromagnetic
fields. We found that the photons propagating on these cosmological geometries
acquire a position-dependent mass-like term (which vanishes when the Hadronic
density vanishes).

Lastly, inflationary models and dark energy theories which follow from the
non-linear sigma model have been widely studied in the literature
\cite{Brown:2017osf,Ema:2020zvg,Paliathanasis:2020wjl}. In the family of
solutions constructed here this possibility is not realized, due to the BPS ansatz.

It is reasonable to expect that the localized configurations we
present in this work are applicable during the early radiation era and right
after quark confinement ($T\sim200MeV$), where the interaction of a radiation
background with the pionic superfluid would be most relevant. In what regards
astrophysical events, we could expect vortices of this type during supernova
expositions which involve huge bursts of hadronic matter. In a future work we
plan to explore further the cosmological and astrophysical implications of our
theoretical construction. Furthermore, we want to explore whether a more
general consideration for the BPS ansatz can provide a mechanism which
explains the early-time expansion of the universe.

\subsection*{Acknowledgements}

F. C. has been funded by Fondecyt grants 1240048 and by Grant ANID EXPLORACION
No. 13250014. The Centro de Estudios Cientificos (CECs) is funded by the
Chilean Government through the Centers of Excellence Base Financing Program of
Conicyt. AG was supported by Proyecto Fondecyt Regular 1240247. AP
acknowledges the support from Fondecyt Regular 1240514. AP gratefully
acknowledges Alex Giacomini and the Universidad Austral de Chile for the
hospitality provided during the completion of this work. AP acknowledges the
COST Action CA23130 ``Bridging high and low energies in search of quantum
gravity (BridgeQG)''.

\end{document}